# Quantum simulation of thermal field theories

**Iván Cuntín[a]*, Wenyang Qian[a], and Bin Wu[a]**

*[a]Instituto Galego de Física de Altas Enerxías, IGFAE, Universidade de Santiago de Compostela*

*E-15782, Santiago de Compostela, Spain*

*E-mail:* ivan.cuntin.broullon@usc.es, qian.wenyang@usc.es, b.wu@cern.ch

*Abstract:* We present our recent studies on thermal field theories using quantum algorithms. We first delve into the representation of quantum fields via qubits on general digital quantum computers alongside the quantum algorithms employed to evaluate thermal properties of generic quantum field theories. Then, we show our numerical results of thermal field theories in 1+1 dimensions using quantum simulators. Both fermion and scalar fields will be discussed. These studies aim to understand thermal fixed points for our forthcoming work on studying thermalisation in quantum field theories in real time quantum simulation.

*Speaker

## 1. Introduction

Quark-gluon plasma (QGP) constitutes one of the main research areas in QCD physics. Time evolution of QCD matter before thermalisation has been studied using classical approaches such as classical field simulations or kinetic theory, but it is yet to be solved from fundamental principles of QCD as a relativistic many-body quantum system. Moreover, lattice QCD is only applicable at low baryon densities where the numerical sign problem does not interfere with calculations. Quantum computing is a potential tool for solving real-time dynamics from QCD first principles.

Quantum computing is a rapidly emerging technology that employs the laws of quantum mechanics to solve problems too complex for classical computers. Currently, we are in the Noisy Intermediate Scale quantum (NISQ) era [1]. Quantum information science has proven valuable in a wide variety of physics applications. These works show that full quantum simulation has the potential to outperform classical methods by reducing problem complexity from exponential to polynomial, which is to be expected based on general arguments relating to local interactions and multi-dimensional Hilbert space [2]. In this work, we use quantum algorithms such as the quantum imaginary time evolution (QITE) algorithm [3] to prepare thermal states and evaluate physical observables in fermionic [4] and scalar [5] thermal field theories.

In order to study the properties of a system in thermal equilibrium we use the formula for the expectation value of an observable:

$$\langle \hat{O} \rangle_\beta = Z_\beta^{-1} \mathrm{Tr}\left[ e^{-\beta \hat{H}} \hat{O} \right],$$

where $\beta = 1/T$ and $Z_\beta = \mathrm{Tr}\left[ e^{-\beta \hat{H}} \right]$ is the partition function. The trace can be calculated by summing the expectation values over the complete set of states. In the cases we are studying, the phase-space distribution $f_p = \langle a_p^\dagger a_p \rangle$ can be readily solved in momentum space.

## 2. Fermion fields in 1+1 dimensions

We start by studying a 1+1 D quantum field theory involving only Majorana fermions with Lagrangian density [3]

$$\mathcal{L} = \frac{1}{2} \bar{\psi} \left( i\gamma^\mu \partial_\mu - m \right) \psi - \mathcal{H}_I(\psi),$$

where $\mathcal{H}_I$ is the interaction Hamiltonian and the fermions satisfy the anti-commutation relations $\left\{ \psi^\alpha(t,x), \psi^\beta(t,y) \right\} = \delta(x-y)\delta^{\alpha\beta}$. The general procedure to represent fields using qubits is as follows [6]: first, we approximate the continuum theory with a discrete theory that's similar to lattice QCD but with continuous time, so that, in the case of free fermions, the Hamiltionian adopts the form:

$$\hat{H} = \frac{1}{2} \sum\nolimits_n \bar{\psi}_n \left[ -\frac{i}{2a} \gamma^1 (\psi_{n+1} - \psi_{n-1}) + m\psi_n \right] - \frac{r}{4a} \sum\nolimits_n \bar{\psi}_n (\psi_{n+1} - 2\psi_n + \psi_{n-1}) + \hat{H}_I,$$

with $\psi_n(t) = \sqrt{a}\psi(t, na)$ and $\hat{H}_I = a \sum \mathcal{H}_I\left( \psi_n / \sqrt{a} \right)$, and where the Wilson term with $r \in (0, 1]$ is included to prevent fermion doubling. Secondly, the fields are mapped into qubits. Following this scheme, we only need $N$ qubits to represent $N$ Majorana fermions.

In coordinate space, we first write the free Hamiltonian in terms of the creation/annihilation operators $a_n^\dagger$ and $a_n$:

$$\hat{H}_0 = \sum_n [-i\frac{a_n a_{n+1} + a_n^\dagger a_{n+1}^\dagger}{2a} + m\left(a_n^\dagger a_n - \frac{1}{2}\right) - \frac{r}{2a}\sum_n [a_n^\dagger(a_{n+1} - 2a_n + a_{n-1}) + 1],$$

and then we do the mapping onto qubits using eigenstates of $a_n^\dagger a_n$ as the computation basis, and employing the Jordan-Wigner transformation

$$a_n^\dagger = \frac{\sigma_n^X - i\sigma_n^Y}{2}\prod_{i=0}^{n-1}\sigma_i^Z \ , a_n = \frac{\sigma_n^X + i\sigma_n^Y}{2}\prod_{i=0}^{n-1}\sigma_i^Z,$$

where $\sigma_n$ are Pauli matrices acting on the n[th] qubit.

This procedure can be carried out in both coordinate and momentum space. However, while simulating free fermion fields in momentum space is more efficient due to the complete diagonalisation of the Hamiltonian, quantum simulations become more efficient in coordinate space once interactions are introduced, as these terms are generally non-local in momentum space. Throughout our work, we carry out simulations in coordinate space, and then compare them with the analytical results derived in momentum space.

In Fig. 1 we show the quantum simulation results for the thermal distribution of the free fields, obtained by evolving the free fermion Hamiltonian with the QITE algorithm. In spite of the limited number of momentum modes, the simulated results of $f_p$ are in strong agreement with the analytical lines, that is, the Fermi-Dirac distributions.

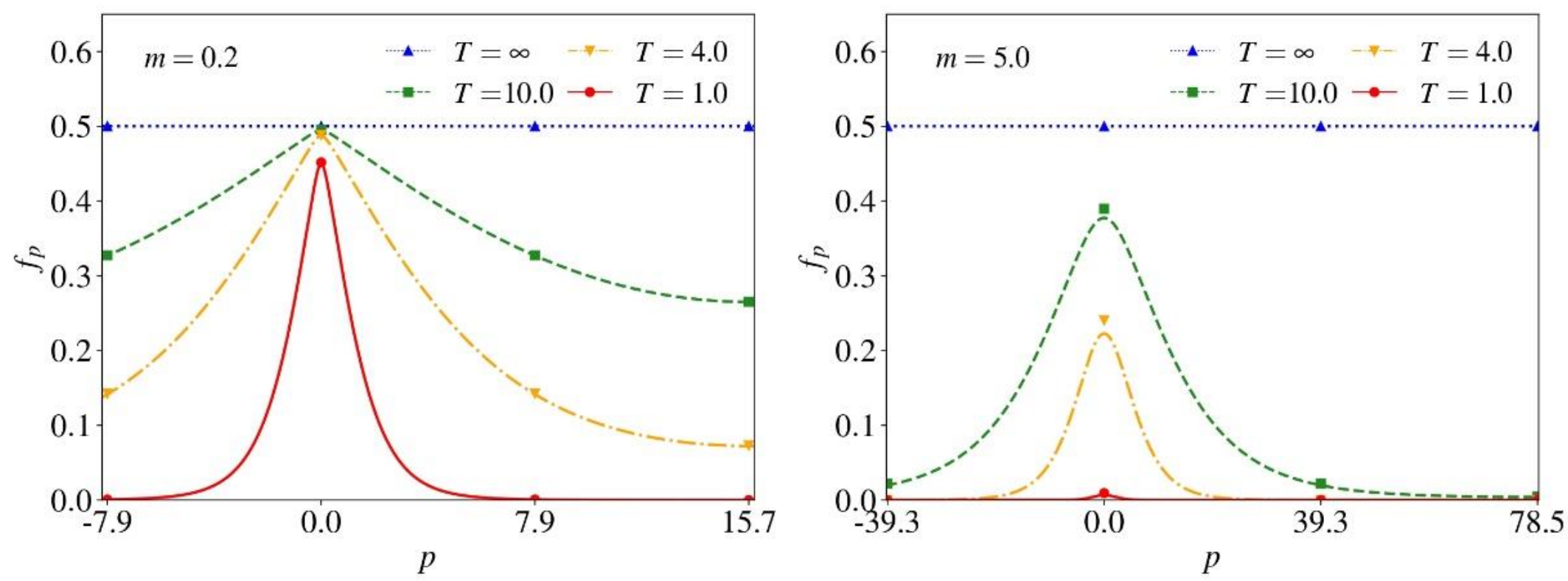

**Figure 1.** Fermionic thermal distribution obtained from quantum simulation on 4 qubits in thermal limits $T \gg m = 0.2$ (left) and $m = 5.0 \gg T$ (right). Simulation results are in solid markers at discretised momenta. Analytical lines of Fermi-Dirac distributions are provided for comparison.

## 3. Interacting fermion fields at thermal equilibrium on qubits

In this section, we investigate fermionic systems coupled through four-fermion interactions. Unlike what happens with Dirac fermions, the four-fermion interactions involving identical Majorana fields effectively vanish due to the Pauli exclusion principle. That means that we need to introduce another Majorana field, $\psi_B = \frac{1}{\sqrt{2}}\begin{pmatrix}1 & 1\\ i & -i\end{pmatrix}\begin{pmatrix}b\\ b^\dagger\end{pmatrix}$, assumed to be homogeneous in space. Accordingly, we have the following expressions for the Lagrangian:

$$\mathcal{L} = \int dx\left[\frac{1}{2}\bar{\psi}(i\gamma^\mu\partial_\mu - m_0)\psi - \frac{g}{4}(\bar{\psi}\psi)(\bar{\psi}_B\psi_B)\right] + \frac{1}{2}\bar{\psi}_B(i\gamma^0\partial_t - M)\psi_B,$$

where $m_0$, $M$ and $g$ are all bare quantities, and for the interaction Hamiltonian:

$$\widehat{H}_I = \frac{M}{2}\bar{\psi}_B\psi_B + \frac{g}{4}\int dx\bar{\psi}\psi\bar{\psi}_B\psi_B.$$

In this model, there's a doubling of the energy levels which gives rise to an additional type of quasiparticle. Accordingly, we define two phase-space distributions:

$$f_p^0 \equiv \langle \hat{a}_p^\dagger \hat{a}_p \rangle_\beta = \frac{Z_\beta^0}{Z_\beta}\frac{1}{1+e^{\beta E_p(\tilde{\mathrm{m}})}}\,, \qquad f_p^1 \equiv \langle \hat{a}'^\dagger_p \hat{a}'_p \rangle_\beta = \frac{Z_\beta^1}{Z_\beta}\frac{1}{1+e^{\beta E_p(\tilde{\mathrm{m}}+g)}}\,,$$

with $\tilde{\mathrm{m}} = E_0 - E_\Omega$ ($E_0$ and $E_\Omega$ being the mass eigenstate and the physical vacuum state, respectively), and two partition functions:

$$Z_\beta^0 = e^{-\beta E_\Omega}\prod_p\left(1+e^{-\beta E_p(\tilde{\mathrm{m}})}\right)\,, \qquad Z_\beta^1 = e^{-\beta E_\Omega^1}\prod_p\left(1+e^{-\beta E_p(\tilde{\mathrm{m}}+g)}\right), \qquad Z_\beta = Z_\beta^0 + Z_\beta^1.$$

Notably, the number operators act on their exclusive subspaces respectively, and they can be rewritten in terms of the fermion fields $a_n^\dagger$ and $a_n$ in practical simulation.

Fig. 2 shows how the results of the previous section are modified by the presence of the background field $\psi_B$. In the simulation we are using $N = 4$ qubits for $\psi$ and one qubit for $\psi_B$ using the QITE algorithm. We see that the simulation results using $a_n^\dagger$ and $a_n$ in position space agree with the analytical calculations using the partition functions.

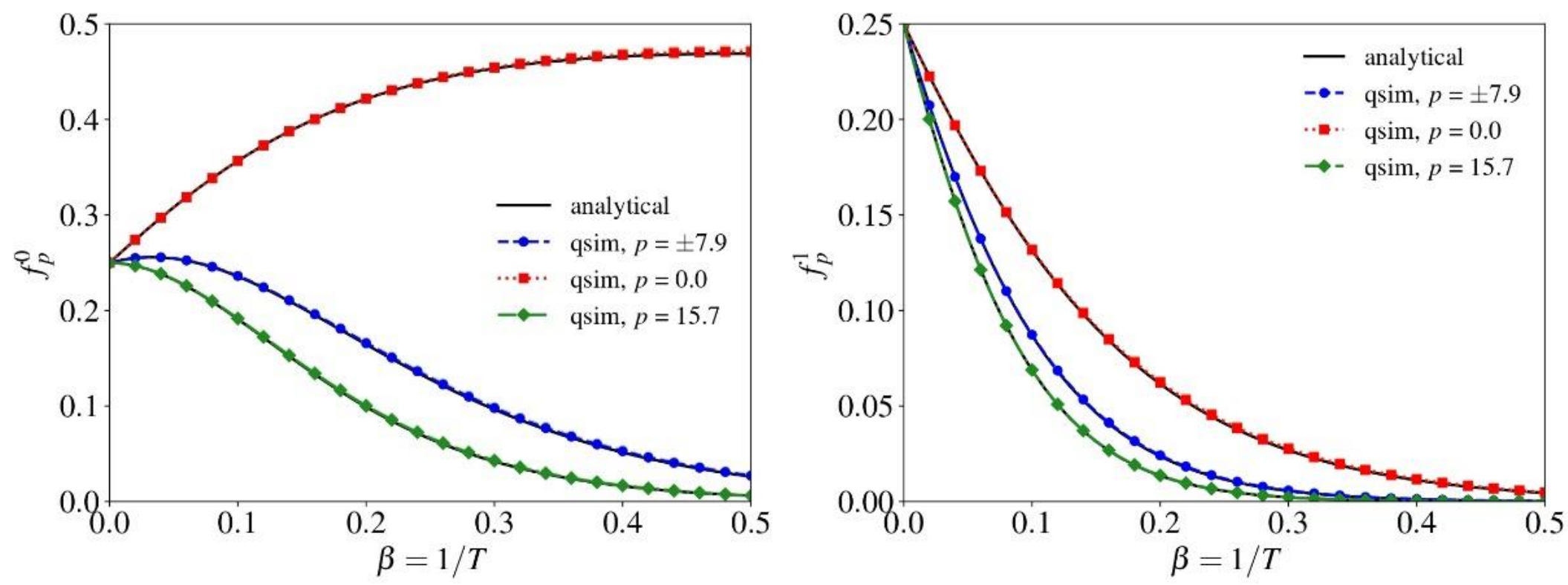

**Figure 2.** Quantum simulation of the thermal distribution functions $f_p^0$ and $f_p^1$ for both quasiparticles.

## 4. Scalar field theory

Moving onto scalar field theory, we'll study quantum simulation as first explored in [7]. The Lagrangian density for the $\phi_4$ theory in (d + 1) dimensions takes the form

$$\mathcal{L} = \frac{1}{2}\left[\partial_\mu\phi\partial^\mu\phi - m\phi^2\right] - \frac{\lambda}{4!}\phi^4.$$

The field operator $\phi$ and its conjugate-field operator $\pi$ satisfy the usual commutation relation $[\phi(x), \pi(y)] = i\delta(x-y)$. Proceeding analogously to the previous section, we start by discretising the hamiltonian, so that it takes the form

$$H_{\mathrm{lat}} = a^d\sum_{n=0}^{N-1}\left[\frac{1}{2}\pi_n^2 + \frac{1}{2}m^2\phi_n^2 + \frac{1}{2}(\nabla\phi)_n^2 + \frac{\lambda}{4!}\phi_n^4\right],$$

where m and λ are the unrenormalised mass and interaction strength. The hamiltonian reduces to

$$\bar{H} = \sum_{n=0}^{N-1}\left[\frac{1}{2}\bar{\Pi}_n^2 + \frac{1}{2}\bar{m}^2\bar{\Phi}_n^2 + \frac{1}{2}(\bar{\Phi}_{n+1} - \bar{\Phi}_n)^2 + \frac{\lambda}{4!}\bar{\Phi}_n^4\right]$$

in 1+1 dimensions and with $\bar{m} = ma$ and dimensionless operators.

Once again, the second step is mapping the bosons into qubits, which can be done in both coordinate and momentum space, although the procedure is more complicated than in the fermionic case, as it will be explained.

Besides having the lattice Hilbert space as a tensor product of local Hilbert spaces, we also have that the local Hilbert space at a single lattice site is infinite dimensional because there are infinitely many bosons contributing to the local wave function, whereas in the case of fermions there were only two. To solve the problem numerically, we truncate the number of bosons by a cutoff number $N_b$ and then digitise the continuous field operators to discretised values.

To do so, we use a finite Hilbert space $\mathcal{H}_n$ of dimension $N_\varphi > N_b$. Then, we define the discrete field operators $\Phi_n$ acting on $\mathcal{H}_n$ as $\Phi_n|\varphi_\alpha\rangle_n = \varphi_\alpha|\varphi_\alpha\rangle_n$ $(\alpha = 0, 1, \dots, N_\varphi - 1)$, where the eigenstates $\{|\varphi_\alpha\rangle_n\}$ are a set of orthonormal vectors in $H_n$ and the eigenvalues are discretised:

$$\varphi_\alpha = \Delta_\varphi\left(\alpha - \frac{N_\varphi - 1}{2}\right) \quad , \quad \Delta_\varphi = \sqrt{\frac{2\pi}{N_\varphi \overline{m}}}.$$

On the other hand, the discrete conjugate-field operators $\Pi_n$ acting on $\mathcal{H}_n$ are constructed via a discrete Fourier transform $\mathcal{F}_n$:

$$\Pi_n = \overline{m}\mathcal{F}_n\Phi_n\mathcal{F}_n^{-1},$$

so that

$$\Pi_n\left|\kappa_\beta\right\rangle_n = \kappa_\beta\left|\kappa_\beta\right\rangle_n \qquad (\beta = 0, 1, \dots, N_\varphi - 1) \quad , \quad \kappa_\beta = \Delta_\kappa\left(\beta - \frac{N_\varphi - 1}{2}\right),$$

with $\Delta_\kappa = \sqrt{2\pi\overline{m}/N_\varphi}$. By replacing the continuous field operators with these new ones, we obtain a discretised, dimensionless scalar field theory with finitely many bosons, and the canonical commutation relation $[\Phi_n, \Pi_n]|n\rangle_n = i|n\rangle_n + \mathcal{O}(\epsilon)$ is satisfied more accurately with increasing dimension of $\mathcal{H}_n$ for lattice sites smaller than the boson cutoff.

To represent the scalar field theory on the qubit, we use a 1D lattice of $N$ quantum registers to represent $N$ lattice points. In each register, we use $n_Q$ qubits so that the Hilbert space dimension is $N_\varphi = 2^{n_Q}$. In this setup, we represent the $\{|\varphi_\alpha\rangle_n\}$ using $n_Q$ qubits on the n[th] quantum registers using binary representation of the label $\alpha$.

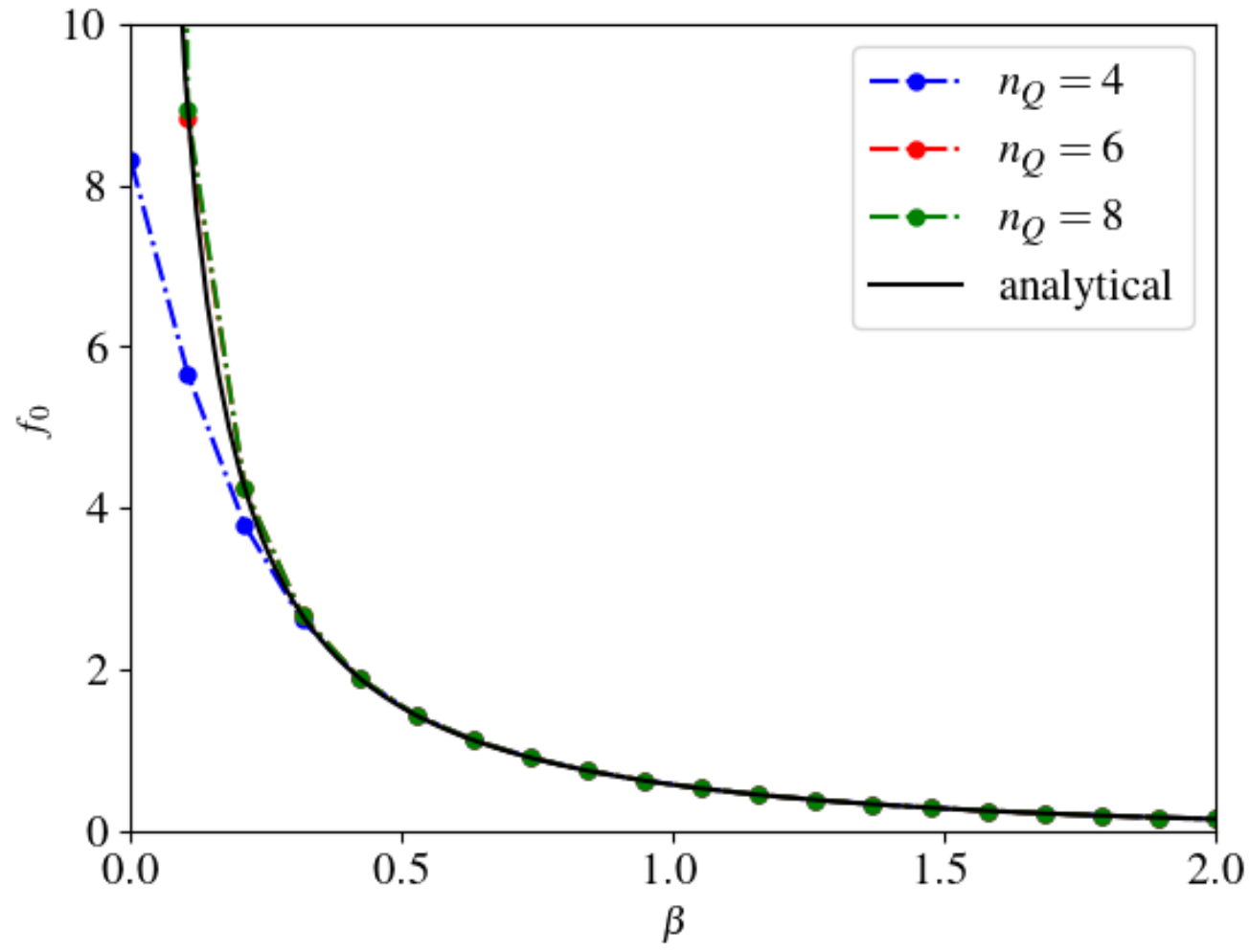

**Figure 3**. Quantum simulation of the thermal distribution as a function of $\beta$.

In Fig. 3, we can see that our numerical results obtained in coordinate space (field operator basis) approach the analytical Bose-Einstein distribution with increasing $n_Q$. This can be straightforwardly generalised to interacting fields.

## 5. Summary and outlook

We develop the QFT for fermionic and scalar fields in 1+1 D using qubits and investigate various thermal properties at finite temperature through quantum simulation algorithms. The numerical results are compared with analytical calculations and exact diagonalisation methods, showing strong agreement. This highlights the potential of quantum computing for field theory computations. Our work represents a significant initial step toward understanding thermal fixed points via quantum simulation. The qubit-based representation effectively yields thermal states towards which a quantum fermion system evolves during the thermalisation process. This lays the groundwork for a more extensive exploration of real-time dynamics in quantum field theory.

## Acknowledgements

This work is supported by the European Research Council under project ERC-2018-ADG-835105 YoctoLHC; by Maria de Maeztu excellence unit grant CEX2023-001318-M and project PID2020-119632GB-I00 funded by MICIU/AEI/10.13039/501100011033; and by ERDF/EU. W. Qian is also supported by the MSCA Fellowships under Grant No. 101109293. B. Wu is also supported by the Ramón y Cajal program with the Grant No. RYC2021-032271-I and the Xunta de Galicia under the ED431F 2023/10 project.